\documentclass[10pt,journal,compsoc]{IEEEtran}
\usepackage[utf8]{inputenc}
\usepackage{array}
\usepackage{xcolor}
\usepackage{enumerate}
\usepackage{amsfonts}
\usepackage{multirow}
\usepackage{graphicx}
\usepackage{url}
\usepackage{bbm}
\usepackage{placeins}

\RequirePackage{fix-cm}
\usepackage{algorithm,eurosym}
\usepackage{algorithmic}
\usepackage{tabulary}



\newtheorem{prop}{Proposition}

\newtheorem{protocol2}{Protocol}
\newtheorem{note1}{Note}

\AtBeginDocument{%
  \providecommand\BibTeX{{%
    \normalfont B\kern-0.5em{\scshape i\kern-0.25em b}\kern-0.8em\TeX}}}

\begin{document}

\title{Secure and Privacy-Preserving Federated Learning via Co-Utility}

\author{Josep Domingo-Ferrer,~\IEEEmembership{Fellow, IEEE,} 
Alberto Blanco-Justicia, Jes\'us Manj\'on and
David S\'anchez%
\thanks{The authors are with the UNESCO Chair in Data Privacy,
CYBERCAT-Center for Cybersecurity Research of Catalonia,
Department of Computer Engineering and Mathematics,
Universitat Rovira i Virgili,
Av. Pa\"{\i}sos Catalans 26, E-43007 Tarragona, Catalonia,
e-mail \{josep.domingo,alberto.blanco,jesus.manjon,david.sanchez\}@urv.cat.
}}%

\markboth{~Vol.~?, No.~?, Month~YYYY}%
{J. Domingo-Ferrer \MakeLowercase{\textit{et al.}}: Secure 
and Privacy-Preserving Federated Learning via Co-Utility} 

\IEEEtitleabstractindextext{%
\begin{abstract}
The decentralized nature of federated learning, that often leverages
the power of edge devices,
makes it vulnerable to attacks against privacy and security.
The privacy risk
for a peer is that 
the model update she computes
on her private data may, when sent to the 
model manager, leak information on those private data.
Even more obvious are security attacks,
 whereby one or several malicious peers return wrong model updates in order
to disrupt the learning process and lead to a wrong
model being learned.
In this paper we build a federated learning framework that offers 
privacy to the participating peers as well as 
security against Byzantine and poisoning attacks.
Our framework consists of several protocols 
that provide strong privacy to the participating 
peers via unlinkable anonymity and that
are rationally sustainable based on the co-utility property.
In other words, no rational party is interested in deviating from
the proposed protocols. 
We leverage the notion of co-utility 
to build a decentralized co-utile reputation management system
that provides incentives for parties to adhere to the protocols.
Unlike privacy protection via differential privacy, our approach
preserves the values of model updates and hence the accuracy
of plain federated learning; unlike privacy protection via
update aggregation, our approach preserves the ability to detect
bad model updates while substantially 
reducing the computational overhead compared to
methods based on homomorphic encryption.
\end{abstract}

\begin{IEEEkeywords}
Federated learning, Model poisoning, Privacy, Security, Co-utility, Peer-to-peer
\end{IEEEkeywords}}

\maketitle

\IEEEpeerreviewmaketitle

\section{Introduction}
\label{introduction}

\IEEEPARstart{F}{ederated} learning~\cite{ref15,mcmahan2017} 
is a decentralized machine learning technique
that allows training a model with the collaboration of multiple
peer devices holding private local data sets that include class labels. 
This approach 
favors privacy because the peers do not need
to upload their private data to a centralized server.
It is also naturally scalable, because 
the computational load is split among the peers, which
may be edge devices such as idle smartphones, and thus widely available. 

In federated learning, a special peer, which we will call the model manager,
sends an initial model to all peers. 
Each peer then computes
a model update by correcting the model so that, 
when input the records in the peer's private data set, 
the model's output 
fits the corresponding class attribute labels.
 Then the peer 
returns the update to the model manager. 
The model manager aggregates the updates and distributes
a new model to the peers. A new learning iteration can now start.
Iterations carry on until the models learned in successive iterations
converge.

Unfortunately, as we discuss in Section~\ref{attacks}, the decentralized nature of federated learning
makes it vulnerable to attacks against privacy and security.
Substantial literature has been devoted to the privacy risks
for peers~\cite{1912.04977}: to what extent the model 
update returned by a peer
can leak her private data.
Privacy-protection techniques include secure aggregation of updates, which
hides individual updates to the model manager, and distortion of updates
via differential privacy, which may significantly hamper the model's accuracy.

Even more obvious are security attacks,
 whereby one or several malicious peers return wrong model updates in order
to prevent the convergence of the model (Byzantine attack) or cause a wrong
model to be learned (poisoning attack).
Protection from Byzantine and poisoning attacks requires the model manager to analyze 
individual peers' updates, thereby making privacy-enhancing techniques based on 
secure aggregation of updates inadequate. 

\subsection*{Contribution and plan of this paper}
\label{contribution}

In this paper we build a federated learning framework that offers both 
privacy to the participating peers and 
security against Byzantine and poisoning attacks.
Our framework consists of several protocols 
designed in such a way that no rational party is interested in acting maliciously.
This makes our protocols robust against security attacks.
Our protocols also provide strong privacy to the participating peers via unlinkable anonymity and without requiring
the aggregation of model updates. In this way, peer updates reach the model manager individually, while
being, at the same time, perfectly accurate. This provides an optimum balance between security, privacy and learning accuracy.

To be rationally sustainable, our protocols are based on the co-utility property~\cite{coutility}.
We also use reputation as a utility to reward well-behaved peers
and punish potential attackers. In order to properly integrate
reputations in the federated learning scenario, our reputation
management is decentralized and itself co-utile.

We report empirical results that show the effectiveness of our protocols at mitigating security attacks 
and at motivating rational peers to refrain from deviating. 

Section~\ref{attacks} discusses privacy and security attacks
against federated learning.
Section~\ref{suite} introduces a co-utile protocol suite
for privacy-preserving and secure federated learning.
Section~\ref{discussion} shows that the proposed protocol
suite achieves co-utility (and hence is rationally sustainable),
privacy and security.
Experimental results are presented in Section~\ref{experiments}.
Finally, conclusions and future research lines are gathered
in~\ref{conclusions}.

\section{Attacks on federated learning privacy and security}
\label{attacks}

In this section, we will discuss the main attacks on privacy
and security that are applicable to federated learning.
For a recent and exhaustive survey, see~\cite{1912.04977}.

\subsection{Privacy attacks}
\label{privacy}


Privacy attacks to federated learning exploit the 
update sent by a peer to infer information on that peer's private 
data.

In federated learning 
the private data sets 
held by the various peers are unlikely to be identically distributed.
What is more, federated learning is explicitly designed 
to improve the learned model by capturing 
the differences among the peers' private
data sets. 
Data inference attacks can be mounted that aim at inferring
how each class is represented in a certain peer's private
data set. 

In~\cite{ref14}, a powerful data inference attack against
federated deep learning is presented that relies on GANs (Generative
Adversarial Networks).
This attack assumes an attacker that can see and use 
internal parameters of the learned model. The attacker participates
as an honest peer in the collaborative learning protocol, but
she tries to extract information about a class of data 
she does not own. 
To that end,
the attacker builds a GAN locally and crafts gradient updates
before returning them in order to influence other participating
peers to leak more information on their data. 
If the attacker is the model manager rather than a peer,
she can do more: the model manager can isolate the shared model
trained by the victim peer. The victim peer's update trained
on the victim's data is used to train the model manager's GAN, that 
can eventually re-create the victim's data.
As explained in~\cite{ref14}, not even differential privacy
used as proposed in~\cite{ref25} can protect against the proposed
GAN attack.

A common requirement of all data inference attacks in federated learning
is that the attacker must be able to link the successive
updates submitted by a certain peer. {\em Our aim is to make
sure that such a linkage is not possible 
by making peers' updates unlinkably anonymous in the model manager's
view.}

\subsection{Security attacks}
\label{security}

Security attacks on federated learning aim at 
disrupting model convergence and thereby 
the learning process. They can be subdivided 
into Byzantine and poisoning attacks.

{\em Byzantine attacks} consist of malicious peers who submit 
defective updates in order to prevent convergence of the global 
model~\cite{ref34}.

Subtler than Byzantine attacks are {\em model poisoning attacks}.
Rather than preventing convergence, the latter aim 
at causing federated learning to converge towards a false global model,
normally one that misclassifies a specific set of inputs.

In~\cite{ref36} it is shown that a single, non-colluding
 malicious peer is enough to mount a poisoning attack.
Yet, security attacks can also be mounted by collusions of peers
or by a single peer masquerading as several peers (Sybil attack).

Countermeasures against Byzantine or poisoning attacks require
seeing the exact values of the individual updates, in order
to assess their goodness.  
This is why some techniques that are good to protect
the privacy of peers, such as secure aggregation of peer updates
via homomorphic encryption~\cite{ref15}, may impair 
the model manager's ability to thwart security attacks.
{\em Our aim is to protect privacy in such a way that malicious
updates can still be attributed.}

\section{A co-utile framework
for privacy-preserving and secure federated learning}
\label{suite}

The foundations of our proposed protocol suite are: i) the notion
of co-utility applied to protocol design and ii) the use of reputations
(computed themselves in a decentralized and co-utile manner) to 
motivate all rational players to behave honestly.
We start by giving some background on co-utility 
and decentralized reputation. Also, for 
convenience Table~\ref{notation}
summarizes the notation used in the rest of this paper.

\begin{table}
\caption{Notation in this paper}
\label{notation}
\begin{center}
\begin{tabular}{|c|m{6cm}|}\hline
Notation & Concept \\ \hline\hline
$M$ & Model manager\\
$AM$ & Accountability manager \\
$P$ & Peer \\
$m$ & Number of $AM$s per peer \\
$\delta$ & Reputation reward/punishment \\
$U$ & Federated learning update \\
$N_U$ & Random nonce encrypted with update $U$\\
$g_i$ & Peer $P_i$'s reputation \\
$PK_M(\cdot)$ & Public key encryption under $M$'s public key\\
$S_{P}(\cdot)$ & Digital signature under $P$'s private key\\
$H(\cdot)$ & Cryptographic one-way hash function\\
$\alpha$ & Flexibility parameter (Note~\ref{notef})\\
$p$ & Forwarding probability\\
$p_0$ & Probability of discarding an update from a peer with zero
reputation\\
$b$ & Size of a batch of non-discarded updates\\
$C$ & Centroid of a batch of updates\\
$T$ & Reputation threshold s.t. if a peer's reputation is at least $T$, her updates are never discarded\\\hline
\end{tabular}
\end{center}
\end{table}

A self-enforcing protocol
is \emph{co-utile}~\cite{coutility} if it results in mutually
beneficial collaboration between the participating agents.
More specifically, a protocol $\Pi$ is co-utile
if and only if {\em the three} following conditions hold:
\begin{enumerate}
\item $\Pi$ is self-enforcing;
\item The utility derived by each agent participating
in $\Pi$ is strictly greater than the utility the agent
would derive from not participating;
\item There is no alternative protocol $\Pi'$
giving greater utilities to all agents
and strictly greater utility to at least one agent.
\end{enumerate}

The first condition ensures that if participants
engage in the protocol, they will not deviate.
The second condition is needed to guarantee that engaging
in the protocol is attractive for everyone.
The third condition can be
rephrased in game-theoretic terms by saying that the
protocol is
a Pareto-optimal solution of the underlying game.

We describe a framework based on co-utility that ensures that
peers can keep their private data sets confidential and, at the same
time, makes them rationally interested in 
returning honest updates to the model manager.

\subsection{Players and security model}
\label{players}

The players in our framework and their security
properties are as follows:
\begin{itemize}
\item {\em Model manager}. The model manager $M$ is a 
player who wants to train a machine learning model 
on the private data of the peers in a peer-to-peer (P2P) network. 
Her interest is to obtain a good quality model, but 
she might be curious to learn as much as possible on the peers'
private data sets. Hence, $M$ can be viewed as {\em rational-but-curious}:
rational to adhere to her prescribed function, but  
curious on the peers' private data.
\item {\em Peers}. They are participants in the network who compute
model updates based on their local private data sets. 
Peers want to preserve their private
data confidential. 
We assume that {\em a majority of peers are rational-but-curious}:
like $M$, they are interested in obtaining
a good quality model, 
 but they also want to influence the model based on their own
respective data;
further, they might be curious to learn as much as possible on the other peers'
private data sets. 
On the other hand, there may be a minority of {\em malicious peers} that
 wish to impair the learning process, because they do not have
the same utility function and/or do not respond to
the same incentives as the rest of peers.
\item {\em Accountability managers}. Accountability managers ($AM$s) 
are randomly chosen peers 
that manage the reputations of other peers. 
Being peers themselves, most accountability managers are rational-but-curious,
but a minority may be malicious.
\end{itemize}

\subsection{Requirements}
\label{requirements}

The assumption that peers are rational rather 
than honest calls
for incentives to make honest behavior attractive to them. 
We will use reputation as an incentive to reward or punish peers. 
In order for this to be effective, the following requirements
need to be fulfilled:
\begin{itemize}
\item {\em Reward.} If a peer contributes a good update, her reputation must increase.
\item {\em Punishment.} If a peer contributes a bad update, her reputation must decrease.
\item {\em Unlinkable anonymity.} Peers contributing good updates 
must stay not only anonymous, but their successive updates must be
unlinkable. 
\item {\em Reputation utility.} Having high reputation
must be attractive for peers. 
Specifically, it must be easier for peers
with higher reputation to contribute their updates while preserving
their privacy. Thus reputation translates to influence without
privacy loss.
\end{itemize}

Unlinkability is our approach to thwarting the privacy
attacks sketched in Section~\ref{privacy}
while perfectly retaining the accuracy of the updates.
On the other hand, reward, punishment 
and reputation utility are our tools to protect
against the security attacks described in Section~\ref{security}.
This will become clear in this section and in Section~\ref{discussion}
below.

\subsection{Co-utile decentralized reputation}
\label{newrep}

Whereas we assume that a majority of peers want to 
learn a good model, we still need to incentivize rational  
peers to abstain from free-riding:
if they find greater utility in deviating from the
federated learning protocol,
they might seriously impair the overall quality of the learned model.
 Also, we need a way to stigmatize/recognize malicious peers in order 
to mitigate their attacks.
To meet the above purposes, we will use reputation management.
In this section we present a reputation management system 
that does not require direct interaction between peers and has
the following interesting properties: 
pseudonymity of peers,
decentralization, 
 resistance
to tampering with reputations,
proper management of new peers 
(to discourage whitewashing bad reputations as new identities
and creating fake peers in Sybil attacks)
and low overhead.

Our reputation protocol maintains a public 
reputation for each peer $P$ that is the result 
of updating $P$'s previous reputation according to the behavior
of $P$ reported by the model manager $M$. Next we explain 
how the above interesting properties are satisfied:
\begin{itemize}
\item {\em Pseudonymity of peers}. Only the pseudonym of 
peers is known, rather than their real identity.
Furthermore, updates that are sent over the network
cannot be linked to the peers that generated them. 
\item {\em Decentralization}. 
The reputation of every peer $P$ is redundantly managed by a number $m$
of peers that act as accountability managers for $P$.
Typically, $m$ is an odd number at least 3 
and the (pseudonymous)
identities of $P$'s
accountability managers are (pseudo)randomly determined by hashing the
peer's pseudonym $P$. In this way, $P$ cannot choose her $m$ accountability
managers, which makes the latter more likely to perform their duty honestly.
\item {\em Tamper resistance}. Since $M$ does not know the 
identity of peers nor is able to link the updates to peers, 
$M$ cannot leverage her position to promote or slander any particular 
peer $M$ likes or dislikes. 
As a consequence, $M$'s rational behavior is to 
exclusively base her reports on the quality of the received model updates.
Regarding tampering by accountability managers, it is thwarted
by their redundancy (see the previous
item on decentralization).
\item {\em Proper management of new peers}. Reputations 
take values in the range $[0,1]$. New peers start with reputation 0,
which makes whitewashing and 
also Sybil attacks unattractive.
\end{itemize}

Let us describe the dynamics
of reputation. Call {\em epoch} the period between
two successive changes of the global model by $M$. 
During an epoch, peers generate and send model updates based
on their private data, with the aim of influencing
the next global model change. Depending on their actions,
peers can earn or lose 
reputation.
Generating a good update increases the generator's reputation
by a certain quantum $\delta/2$ fixed by the model manager;
furthermore,
helping a good update reach the model manager
in a way unlinkable to the generator brings a $\delta/2$
reputation increase to one of the helping peers. Thus, 
every good update results in a total $\delta$ reputation 
increase.
On the other hand, generating a bad update decreases the generator's reputation
by $\delta$. Thus, the overall reward for a good update equals
the punishment for a bad update.

Some peer reputations may become negative and some may become
greater than 1 as an epoch progresses.
At the epoch's end, reputations are re-normalized
into the range $[0,1]$ as follows. First, accountability
managers reset any negative reputation to 0.
Then, if there are reputations above 1, 
 all reputations are divided by 
the largest reputation. To that end, when a peer's reputation
becomes larger than 1, the peer's accountability managers
broadcast that reputation, which allows all accountability 
managers to compute the maximum reputation reached in that epoch
and thereby normalize all reputations into the interval $[0,1]$.


Normalization has the beneficial effect of 
deterring free-riding: even if a peer
has attained high reputation, she will lose it 
gradually if she stops participating.
 Indeed, any peer's reputation will decrease 
due to normalization unless
she continues to generate good updates or helps routing them.    
This addresses the second condition of the co-utility definition:
the utility derived from participating must be greater than the utility
derived from not participating.
Fulfillment of the other two conditions for co-utility
will be justified in Section~\ref{dis-coutil} below.



\subsection{Downstream: from update generator to model manager}
\label{down}

We call downstream operation the submission of 
model updates from the peers to the
model manager $M$. In order to preserve privacy and encourage
security, we propose Protocol~\ref{pro1}. In 
Section~\ref{discussion}, we will show that it is co-utile.

The idea of Protocol~\ref{pro1} is that a peer, say $P_1$, 
 does not directly
send her update to $M$. Rather, $P_1$ asks another peer, say 
$P_2$, to do so. $P_2$ randomly decides 
whether to submit $P_1$'s update to $M$
or forward it to another peer, say $P_3$, who stands the same
choice as $P_2$. Forwarding continues until a peer
is found that submits the update to $M$.
\vspace{1ex}
\begin{protocol2}[Update submission]
\label{pro1}
\begin{enumerate}
\item\label{pas1} Let $P_1$ be a peer that generates an update $U$.
Then $P_1$ 
encrypts $U$ along with a random nonce $N_U$ under the model manager's
public key, to obtain $PK_M(U,N_U)$ (we assume the message $U,N_U$ to 
have a certain format that allows distinguishing it from gibberish at
decryption).
In this way, only $M$ will be able to recover the update $U$.
The generator $P_1$ {\em never} submits her own update to the manager $M$;
rather, $P_1$ forwards $S_{P_1}(PK_M(U,N_U),H(H(H(U,N_U))),P_2)$,
where $H$ is a one-way hash function and $S_{P_1}$ is $P_1$'s signature,
to another peer $P_2=\mbox{\sc Select}(g_1)$, where function 
{\sc Select}() is explained below.
\item\label{pas2} If $P_1$'s reputation $g_1$ is such that 
$g_1 < \min(g_2,T) - \alpha$,
where $g_2$ is $P_2$'s reputation,
$T$ is a parameter such that updates submitted by peers
with reputation $T$ or above are never discarded,
 and $\alpha$ is a flexibility
parameter discussed in Note~\ref{notef}, 
then $P_2$ discards the received update.
Otherwise, 
$P_2$ makes a random choice: with probability $1-p$, she submits
\[S_{P_2}(PK_M(U,N_U), H(H(H(U,N_U))), M)\]
to $M$ and with probability $p$ she forwards
\[S_{P_2}(PK_M(U,N_U), H(H(H(U,N_U))),  P_3)\]
to another peer $P_3=\mbox{\sc Select}(g_2)$.
\item\label{pas3} 
If $P_2$'s reputation is below $\min(g_3,T) - \alpha$ 
 then 
$P_3$ discards the received update.
Otherwise, 
$P_3$ makes a random decision as to submit or forward.
If it is forward, $P_3$ will use the {\sc Select}() function
and there may be more peers involved: $P_4$, $P_5$, etc.
\item\label{pas4} Eventually $M$ receives an update 
\[S_{P_i}(PK_M(U,N_U), \allowbreak H(H(H(U,N_U))), M)\] 
from a peer $P_i$. Upon this, $M$ does:
\begin{enumerate}
\item\label{pasdiscard}
 Directly discard the update with probability
$p_0 (1-\min(g_i/T, 1))$,
where $p_0$ is a parameter
indicating the probability of discarding an update
submitted by a peer with 0 reputation, and $g_i$ is $P_i$'s 
reputation.
\item\label{pas4a} If the update 
has not been discarded, decrypt $PK_M(U,N_U)$, obtain $U$, check 
that the nonce $N_U$ was not received before
(to make sure $U$ is not a replay of a previously received update)
 and check the hash $H(H(H(U,N_U)))$.
\item\label{pas4b}  Wait until a batch of $b$ non-discarded updates
has been received in order to be able to decide
whether $U$ is good or bad 
(see Section~\ref{securitysec} below on how to detect
bad updates). 
\item\label{pas4c} Change 
the model with the good updates in the batch and
publish the updated model.
\item\label{pas4dstar} Publish the value $\delta=1/b$.
\item\label{pas4d} For every good non-discarded update $U$, publish
$H(H(H(U,N_U)))$.
\item\label{pas4e}  For every bad non-discarded 
update $U$, call {\sc Punish}($P_i$)
where $P_i$ is the peer having submitted $U$
and {\sc Punish}() is Protocol~\ref{pro2} in Section~\ref{upstream}.
\end{enumerate}
\end{enumerate}
\end{protocol2}

Function {\sc Select}($g_i$) is used by a peer $P_i$ to select
a forwardee. There are several ways in which this can be accomplished.
However, the rational choice is for $P_i$ to select
a forwardee $P_j$ with a sufficient reputation so that 
$M$ does not reject the update should $P_j$ submit it directly
to $M$. Hence, if $P_i$'s reputation is $g_i \geq T - \alpha$,
$P_i$ can randomly pick any of the peers whose reputation
is $T$ or above, because none of those peers risks update discarding.
However, if $g_i < T - \alpha$, $P_i$ chooses the peer
with the maximum reputation that does not exceed
$g_i + \alpha$, because no 
peer with reputation above that value will accept to
forward $P_i$'s update.


\begin{note1}[On the flexibility parameter $\alpha$] 
\label{notef}
In Protocol~\ref{pro1} a peer accepts to forward updates
from  peers that have at least her own
reputation minus a flexibility amount $\alpha$. Using a small
value $\alpha >0$ introduces some flexibility and helps
new peers (that start with 0 reputation) to earn 
reputation as generators or first forwardees of good updates.
Large values of $\alpha$ are not acceptable from the rational
point of view: high-reputation peers have little to gain by
accepting updates from peers who 
are much below them in reputation, because the latter
are likelier to convey bad updates or to fail to reward
the first forwardee in case of good updates.
\end{note1}

\begin{note1}[On loops, multiple paths and other misbehaviors]
Nothing is gained by any peer if loops arise accidentally or intentionally
in Protocol~\ref{pro1}.
As it will be seen in below (Protocol~\ref{pro3} and Note~\ref{1streward})
only the first peer chosen by the update generator 
is rewarded. Hence, forwarding twice or more
times the same message brings no additional benefit. On the other hand,
a generator $P$ might send the same good update through several
paths to increase the reputation of several first peers. However,
by promoting more peers than necessary, $P$ may experience a decrease
of her own reputation, because reputations are normalized when
any peer reaches a reputation above 1 (see Section~\ref{newrep}). 
Finally, update generators could systematically choose themselves as first
forwardees of good updates 
to collect additional reward; but if they do so, they weaken their privacy.
\end{note1}

\begin{note1}[Key generation]
In Protocol~\ref{pro1}, peers sign the messages they send. To that
end, each peer needs a public-private key pair. At least the two following 
alternative key generation procedures are conceivable:
i) identity-based signatures, in which the peer's pseudonym
is her public key and the peer's private key is generated
by a trusted third-party~\cite{ident}; ii) blockchain-style 
key generation~\cite{blockchain}, 
in which the peer generates her own 
key pair without the intervention of any trusted third-party
or certification authority, and then obtains her pseudonym
$P_i$ (her {\em address} in the blockchain network) as a function
of her public key.     
\end{note1}

\subsection{Upstream: from model manager to update generator}
\label{upstream}

By upstream operation we denote the punishment of bad updates
and the reward of good updates. Let us start with Protocol~\ref{pro2}
that seeks to penalize the generator of a bad update by retracing
the reverse path from $M$ to the generator.
The peer $P_i$ who submits an
update found to be bad by the manager can escape punishment 
if $P_i$ can show to her accountability managers that she received
the bad update from a previous peer, say $P_{i-1}$.

\vspace{1ex}
\begin{protocol2}[{\sc Punish}($P_i$)]~\\
\label{pro2}
Every accountability manager $AM$ of $P_i$'s does:
\begin{enumerate}
\item Ask $P_i$ whether $P_i$ can prove she did not generate $U$.
\item If $P_i$ can show to $AM$ a message
\[ S_{P_{i-1}}(PK_M(U,N_U),H(H(H(U,N_U))),P_i) \]
then 
\begin{enumerate}
\item Do not punish $P_i$ (the peer's reputation is left intact);
\item Call {\sc Punish}($P_{i-1}$).
\end{enumerate}
Otherwise, punish $P_i$ by decreasing her reputation by $\delta$.
\end{enumerate}
\end{protocol2}

The punishment protocol must be initiated by $M$, because the 
model manager is the
only party that can detect bad updates and that is interested 
in punishing them. However, the punishment is actually executed
by the guilty peer's accountability managers. Hence, $M$ cannot
track which peer is actually punished for that bad update, which
prevents $M$ from identifying the generator of an update 
by (falsely) claiming that the update is bad.

Unlike the punishment protocol, 
the rewarding protocol is initiated by the peer who submitted
a good update, because that peer is the one interested in the reward.
As we will later justify, the first peer (and only
the first peer) who is asked by 
the generator to submit or forward a good update is also entitled
to a reward. We will call that peer the ``first forwardee''.

\vspace{1ex}
\begin{protocol2}[{\sc Reward}($U$)]
\label{pro3}
\begin{enumerate}
\item\label{reward1} When $M$ publishes $H(H(H(U,N_U)))$ for a good update, then
the update generator, say $P_1$, sends to the first forwardee, say $P_2$, 
$S_{P_1}(H(H(U,N_U)),P_2)$.
\item\label{reward2} $P_2$ 
checks that the hash of $H(H(U,N_U))$ matches\\
$H(H(H(U,N_U)))$ published by $M$. If it is so, $P_2$
returns a receipt $S_{P_2}(H(H(U,N_U)), P_1)$ to the generator $P_1$.
\item\label{reward3} $P_1$ proves to her
accountability managers that she is the generator by 
showing $H(U,N_U)$ to them and proves that she has acknowledged her  
first forwardee by showing the receipt $S_{P_2}(H(H(U,N_U)), P_1)$.
\item\label{reward4} Every accountability manager $AM$ of $P_1$'s 
checks $P_2$'s receipt and checks that 
the {\em double} hash of $H(U,N_U)$ received from $P_1$ matches
$H(H(H(U,N_U)))$ published by $M$. If {\em both} checks are fine, 
$AM$ increases $P_1$'s reputation by $\delta/2$.
\item\label{reward5} $P_2$ sends
 $S_{P_1}(H(H(U,N_U)),P_2)$ to 
her accountability managers to claim her reward. 
\item\label{reward6} Every accountability manager $AM$ of $P_2$'s 
checks that the hash of $H(H(U,N_U))$ matches 
$H(H(H(U,N_U)))$ published by $M$. If it is so,
$AM$ increases $P_2$'s reputation by $\delta/2$.
\end{enumerate}
\end{protocol2}

\begin{note1}[On rewarding the first forwardee only]\label{1streward} In Protocol~\ref{pro3} only
the first forwardee is rewarded, rather than all forwardees. 
The reason is that we want the total budget to reward a good update
to be fixed and equal to the budget $\delta$ used to punish a bad update.
We also want the reward share for the generator of a good update to 
be fixed, say $\delta/2$, and independent of the number of hops
the update travels before reaching $M$.
Hence, if we chose to reward all forwardees, the fixed reward 
share $\delta/2$ for forwardees ought
to be distributed among them.
Therefore, every forwardee would be 
better off by submitting the update to $M$ rather than forwarding it
to another forwardee who would take part of the reward.
As a consequence, there would be only one forwardee, who would know 
that the previous peer is the generator of $U$. This would break privacy.
Rewarding only the first forwardee avoids this problem 
and is a sufficient incentive,
because any forwardee can hope to be the first (due
to the protocol design, a forwardee does not know
whether she receives an update from the generator
or from another forwardee) and thus has a reason
to collaborate.
\end{note1}

\begin{note1}[On peer dropout]
Accidental (due to power or network failure) or intentional peer dropout
does not affect the learning process: on the one hand, once an update
has been generated/forwarded, the generator/forwarder can disappear;
on the other hand, the next forwardee is chosen among the peers
who are online. Reputation management is also resistant to dropout
of accountability managers, because there are $m$ of them for each peer; 
$m$ just needs to be increased if dropout is very likely. Punishment
is not affected: even though a peer drops out, he will be punished
with a reputation decrease all the same. However, rewarding may be 
problematic in the very specific case
that either the update generator $P_1$ or the first 
forwardee $P_2$ drop out before rewarding is complete: 
the one of the two that remains online
may not receive her/his reward. 
\end{note1}

\section{Discussion}
\label{discussion}


In this section, we first demonstrate that the framework formed
by Protocols~\ref{pro1},~\ref{pro2} and~\ref{pro3} is co-utile, 
that is, that those protocols will be adhered to by the players
defined in Section~\ref{players}.
Then we will show that the protocols satisfy the 
requirements of Section~\ref{requirements}, and thereby
preserve the confidentiality of the users' private data
and protect the learned model from Byzantine and poisoning 
security attacks.

\subsection{Co-utility}
\label{dis-coutil}

To argue co-utility for  
 Protocols~\ref{pro1}, \ref{pro2} and~\ref{pro3},
 we must show that following them is a better
option for $M$ and the peers than deviating.

\subsubsection{Co-utility for the model manager}

The model manager's goal is to train a model based
on the peers' private data sets. For that reason,
$M$ is interested in encouraging good updates and punishing
bad updates. On the other hand, $M$'s role is limited 
to Step~\ref{pas4} of Protocol~\ref{pro1}.
Let us examine in detail the actions of $M$ in that
step and whether $M$ could gain by deviating from them
or skipping them:
\begin{enumerate}
\item In Step~\ref{pasdiscard}~$M$ directly discards an update
with a probability that is inversely proportional to the reputation
of the submitting peer. Discarding is only based on reputation,
without examining whether the update is an outlier. $M$ is 
interested to perform this step at least for two reasons: 
first, it reduces $M$'s computational overhead,
and second, it allows $M$ to make 
reputation attractive for peers (only high-reputation
peers, those with reputation at least $T$, are sure of getting
their updates examined). At the same time, if $M$ wants
 to keep the peer community alive, $M$ should allow
a nonzero probability $1-p_0$ 
of examining an update submitted by a new peer (that has 0 reputation). 
Also, setting up a threshold $T$ above which updates are examined
for sure is a way for $M$ of not losing too many good updates.
\item Step~\ref{pas4a} consists of decrypting the update,
checking its freshness and checking that the hash is correct.
Obviously, $M$ is interested in carrying out these steps.
Without the updates, $M$ cannot train the model.
\item Step~\ref{pas4b} is about deciding whether an 
update is good or bad. $M$ clearly needs to make this decision,
in order to use good updates to improve the model and punish
bad updates to discourage them.
\item Step~\ref{pas4c} is about changing the model using the good
updates. This is exactly $M$'s main goal.
\item Step~\ref{pas4dstar} publishes $\delta$ that determines
the amount whereby 
reputations must be increased/decreased by the accountability managers.
$M$ is interested in publishing $\delta$ to 
facilitate a correct reputation management that
 keeps peers incentivized. In fact, if the number $b$ of updates
per batch is fixed, then $\delta$ is also fixed and 
 does not need to be published at each protocol execution.
\item Step~\ref{pas4d} publishes information that peers can 
use to claim rewards for good updates. If $M$ deviates and does not publish
this information, then peers cannot claim rewards. This would discourage
peers from submitting good updates and would be against $M$'s interests.
\item Step~\ref{pas4e} launches the punishment procedure for each
bad update. If $M$ did not perform this step, bad updates would go 
unpunished, which would fail to discourage them.
\end{enumerate}

\subsubsection{Co-utility for the update generator}

In Protocol~\ref{pro1}, the update generator only works in Step~\ref{pas1}.
Let us analyze the actions in this step:
\begin{enumerate}
\item {\em Update generation and encryption}.
The generator, say $P_1$, generates an update
and encrypts it together with a random nonce so that only $M$
can decrypt the update and check its freshness:
\begin{enumerate} 
\item The intrinsic motivation for $P_1$ to generate an update 
is to have an influence on the model being learned: a rational peer
wants to help obtain an accurate model that is socially beneficial
in some sense, whereas a malicious peer wants to 
poison the learned model. 
\item The motivation for $P_1$ to generate a {\em good} update
$U$ is to keep her reputation high. A high reputation brings more 
influence on the model learning.
Specifically, a high $g_1$ allows $P_1$ to find $P_2$ 
 such that $g_1 \geq g_2 - \alpha$, which means that $P_2$ does not 
discard $P_1$'s update, and with $g_2$ high enough 
for $P_1$ to be confident that $P_2$ can be entrusted with 
relaying $U$ towards $M$ with little or no probability of 
$U$ being discarded by $M$ without examination (see
description of the {\sc Select}() function in Section~\ref{down}).
If $U$ eventually reaches $M$, this brings $P_1$ influence and further
reputation increase, which means more influence in the future.
\item The motivation for $P_1$ to encrypt $U$ under $M$'s public 
key is to prevent
anyone else from claiming the reward for that update, should $U$ be good.
The motivation for $P_1$ to sign the forwarded message is that 
 the forwardee $P_2$ will not accept an unsigned message, because
$P_2$ will need that signed message to escape punishment in case
 $U$ is bad.
\end{enumerate}
\item {\em Update forwarding}. In terms of privacy, 
it is bad for $P_1$ to submit her generated update directly to $M$,
as it could leak information on her private data set. It is still bad
 if $P_1$ directly submits with probability $1-p$ and 
forwards with probability $p$, like in the Crowds system~\cite{crowds}. 
If we used the Crowds algorithm, 
from the point of view of $M$ the most likely submitter
of an update would be the update generator: $U$ would be submitted by 
$P_1$ with probability $1-p$, whereas it would be submitted by
the $i$-th forwardee with probability $(1-p) p^i < 1-p$.
Hence, $P_1$ is interested in looking for a forwardee $P_2$ who
takes care of her update, rather than 
submitting her update herself. Specifically, $P_1$ wants a forwardee
$P_2$ such that: a) $P_2$ will accept to forward
$P_1$'s update; b) $P_2$ does not risk update discarding 
($g_2 \geq T$) or risks it 
with the smallest possible probability (see 
the description of the {\sc Select}() function in Section~\ref{down}).
Further, if $P_1$ can choose among several possible $P_2$ with 
$g_2 \geq T$, $P_1$'s best option is to pick $P_2$ randomly
for the sake of unlinkability of successive updates to each other.
Here we see a second benefit of a high reputation for $P_1$:
 the higher $g_1$, the more peers with reputation at least $T$
$P_1$ can choose from and the higher is unlinkability.
\end{enumerate}

In Protocol~\ref{pro2}, the update generator $P_1$ has a role only if her 
update is bad. The generator's role in this case is a passive 
and inescapable one:
when $P_1$ is asked by her accountability managers 
to show that $P_1$ received the bad update from someone else, 
$P_1$ cannot show it and is punished. 

In Protocol~\ref{pro3}, the generator $P_1$ of a good update 
is clearly interested in running Step~\ref{reward1} of the 
protocol to claim a reward. 
In Step~\ref{reward1}, $P_1$ is forced to give the first forwardee
$P_2$ the necessary information $H(H(U,N_U))$ so that $P_2$ can 
claim his reward. The reason is that, without $P_2$'s receipt,
$P_1$ cannot claim her own reward at Step~\ref{reward3} (this 
latter step is also self-enforcing if $P_1$ wants her reward).

$P_1$ could certainly decide to favor a false first forwardee
$P'_2$ of her choice, rather than the real first forwardee $P_2$.
This would still work well for $P_1$, because 
$P'_2$ would return a signed receipt for the same reasons
that $P_2$ would do it.
However, if $P_1$ wants to favor $P'_2$, it entails
less risk (of being discovered) 
for $P_1$ to use $P'_2$ as a {\em real} first forwardee.
Thus, there is no rational incentive to favor
false first forwardees.

\subsubsection{Co-utility for the update forwardees}

In Protocol~\ref{pro1}, the forwardees $P_2, P_3, \ldots$ work in 
Steps~\ref{pas2} and~\ref{pas3}, which are analogous to each other.  
Let us examine the actions expected from a forwardee:
\begin{enumerate}
\item {\em Update acceptance or discarding}. The incentive
for a forwardee $P_i$ to accept to deal with an update $U$ 
is to be rewarded in case $U$ is good and $P_i$ is the first
forwardee (note that $P_i$ does not know whether she is the first,
but hopes to be). 
Thus, if $P_i$ receives
the update from a previous peer $P_{i-1}$ with high reputation, 
$P_i$'s rational decision is to accept that update: 
there are chances that $U$ is good, which will bring reward
if $P_i$ turns out to be the first forwardee. 
In contrast, if $U$ comes from a peer $P_{i-1}$ with low reputation,
it is less likely that the update is good, so $P_i$'s rational
decision is to discard $U$ to avoid working and 
spending bandwidth for nothing.
\item {\em Update submission or forwarding}. 
It takes about the same effort for a forwardee $P_i$ to submit
an update to $M$ or to forward it to some other peer $P_{i+1}$.
Hence, it is rational for $P_i$ to make the decision randomly according
to the prescribed probabilities ($1-p$ for submission and $p$ for 
forwarding). In case of forwarding, $P_i$'s rational procedure
is like the generator's: look for a forwardee 
with reputation at least $T$ if $g_i \geq T - \alpha$ or
the maximum possible reputation that 
does not exceed $g_i + \alpha$ otherwise (as per the
{\sc Select}() function.
Also, no matter whether forwarding or submitting, $P_i$ has to replace
the previous signature of the update by her own signature: neither
the model manager nor any forwardee will accept from $P_i$ a message
that is not signed by $P_i$, because they will need the signed message
in case $U$ turns out to be bad and punishment is launched.
\end{enumerate} 

In Protocol~\ref{pro2}, if $P_i$ did not generate a bad 
update $U$, $P_i$ will rationally do her best to avoid punishment (reputation decrease)
by showing a message signed by whoever sent $U$ to her. 

In Protocol~\ref{pro3}, $P_2$'s best option is to return the receipt
at Step~\ref{reward2}, because $P_1$ could otherwise blacklist $P_2$
and never make $P_2$ a first forwardee in future epochs.
Finally, $P_2$ is obviously interested in claiming her reward
in Step~\ref{reward5}.

\subsubsection{Co-utility for the accountability managers}

The accountability managers are a keystone in Protocols~\ref{pro1},
\ref{pro2} and~\ref{pro3}. In our security model (Section~\ref{players})
a majority of them is assumed to be rational and to be 
 interested in obtaining a well-trained model. 
Hence, a majority
of the $m$  
accountability managers pseudorandomly assigned to 
each peer can be expected to behave honestly,
which in turn means that the reputation  
of every peer can be expected to be honestly managed.

In Protocol~\ref{pro1}, there is no direct intervention of 
accountability managers. It suffices that they honestly maintain 
and supply
the reputations $g_i$ of all involved peers $P_i$ as described
in Section~\ref{newrep}.

As to Protocol~\ref{pro2}, it is launched at the request
of $M$ in the last step of Protocol~\ref{pro1}. 
In Protocol~\ref{pro2}, the accountability managers have the lead role.
Most of each peer's accountability managers 
can be assumed rational and therefore 
they can be assumed to discharge their role as described in the protocol.

Finally, in Protocol~\ref{pro3}, the 
 accountability managers of the generator reward the latter 
in Step~\ref{reward4}.
Then in Step~\ref{reward6} the first forwardee is 
rewarded by her  accountability managers. 
Again, since for each peer a majority
of accountability managers can be assumed rational, we can expect
them to honestly perform those two steps as described in Protocol~\ref{pro3}. 

\begin{note1}[Non-collusion scenario]
\label{noncoll}
In fact, given that the accountability managers assigned to a peer
are randomly chosen, it is reasonable to assume that in general
they do not know each other and hence they do not collude.
 In the non-collusion scenario, not even a majority 
of honest accountability managers is needed. 
If malicious accountability managers do not collude, 
each of them is likely to report different reputation results.
Hence, as long as {\em two} of the peer's accountability managers act
rationally and follow the protocol, their correct
result is likely to be the most frequent one and thus to prevail.
\end{note1}

\subsection{Privacy}

As mentioned in Section~\ref{privacy}, ensuring the unlinkability
of updates goes a long way towards guaranteeing that the 
private data sets of peers stay confidential.
We can state the following proposition.

\begin{prop} 
\label{proppriv}
If 
the forwarding probability is $p>0$ and
there is no collusion between the model manager $M$ and 
peers, the private data set of each peer remains 
confidential versus the model manager and the other peers.
Confidentiality is based on update encryption and unlinkability,
and unlinkability increases with $p$ and the generator's reputation.
\end{prop}

\begin{IEEEproof} The privacy guarantee is based on unlinkability
and update encryption.

Let us first consider linkability by $M$.
By the design of Protocol~\ref{pro1}, $M$ knows that 
the submitter of an update $U$ is never the update generator.
At best, $M$ knows that the probability 
that $U$ was submitted by the $i$-th forwardee is $(1-p)p^{i-1}$,
and hence that the most likely submitter is the first forwardee.
However:
\begin{itemize}
\item 
The larger $p$, the greater the uncertainty about 
the number of hops before the update is submitted,
and hence the harder for $M$ to link a received update
to its generator.
\item The next forwardee is selected using the {\sc Select}()
function, described in Section~\ref{down}.
If $g_{gen} \geq T - \alpha$, then $P_{gen}$ chooses 
the first forwardee randomly
among the set of peers with reputation at least $T$, and this 
set depends on the current reputations and varies over time; 
hence, as long as there are several peers with reputation 
$T$ or above, the fact that two updates were submitted
by the same peer does not tell $M$ that both updates were
generated by the same peer.
If $g_{gen} < T - \alpha$, then $P_{gen}$ chooses as a first forwardee 
the peer with the maximum reputation that does not exceed
$g_i + \alpha$: if reputations do not change between two successive
updates, $P_{gen}$ would choose the same first forwardee 
for both updates; yet, $M$ cannot be sure that the submitter
of both updates is really the first forwardee, and hence 
$M$ cannot be sure that both updates were generated
by the same $P_{gen}$. Hence, in no case 
can two different updates by the same generator be
unequivocally linked, even if the probability of correctly
linking them is lower when $g_{gen} \geq T - \alpha$.              
\end{itemize}

On the other hand, {\em neither
the reward nor the punish protocols allow $M$ to learn
who generated a good or a bad update}.
Thus, $M$ can neither link the updates he receives nor 
unequivocally learn
who generated a certain update $U$.
Therefore, $M$ cannot obtain any information on the private
data set of any specific peer $P$.


Consider now linkability by a peer $P_i$:
\begin{itemize}
\item If $P_i$ is a forwardee for 
two different updates from $P_{i-1}$ 
and $p>0$, $P_i$ does not
know whether $P_{i-1}$ generated any of the updates or is merely
forwarding them.
$P_i$'s uncertainty about $P_{i-1}$ being the generator 
is Shannon's entropy $H(p)$, which grows with $p$ for $p\leq 0.5$;
for $p>0.5$, what grows with $p$ is $P_i$'s certainty that 
$P_{i-1}$ is {\em not} the generator. In summary, $P_i$ can only
guess right that $P_{i-1}$ is the generator if $p$ is very small: 
in this case, forwarding hops after the first mandatory
hop from generator to first forwardee are very unlikely.
\item The only exception is when $P_i$
is the first forwardee for two good updates from the
same generator (because in this case he receives a message from
the generator in Step~\ref{reward1} of Protocol~\ref{pro3}).
However, in this case $P_i$ can only link the {\em encrypted} 
version of updates (that is, $PK_M(U,N_U)$ and $PK_M(U',N_{U'})$),
but has no access to the clear updates $U$, $U'$. Hence, $P_i$
gets no information on $P_{i-1}$'s private data set.
\item If $P_i$ is an accountability manager of a generator $P_j$,
$P_i$ can link all {\em encrypted} good updates originated by 
$P_j$. However, since those updates are not in the clear,
$P_i$ gets no information on $P_j$'s private data set.
\end{itemize}
\end{IEEEproof}

Note that assuming there are no collusions is plausible
because peers are pseudonymous: normally people collude only
with those they know. 

A successful collusion must include one or more first forwardees
(who know the pseudonyms of the update generators) and 
$M$ (who can decrypt the updates).
 In this way, $M$ can attribute updates 
and perhaps link those corresponding
to the same generator; then $M$ can infer whatever information
on the generator's private data set is leaked by the generator's updates.

However, to allow update linkage, a collusion requires a malicious model
manager and  a significant proportion  
of malicious peers, whereas 
in our security model (Section~\ref{players}) we assume
$M$ and a majority of peers to be rational-but-curious.
A collusion of $M$ with a substantial number of peers
is hard to keep in secret, and if it becomes known that $M$ is 
 malicious, peers will be unwilling 
to help $M$ to train the global model. Therefore, $M$'s rational
behavior is to abstain from collusion.

\subsection{Security}
\label{securitysec}

Guaranteeing security means thwarting  Byzantine and poisoning attacks
(Section~\ref{security}), which consist of submitting bad model updates. 
We first recall the approaches that have been proposed in
the federated learning literature for the model manager
to defend against bad updates.
They fall into the following three broad classes
(see the surveys~\cite{1912.04977,blanco} for more details):
\begin{itemize}
\item {\em Detection via model metrics.} An update is
labeled as bad if incorporating it to the model degrades
the model accuracy. This approach requires a validation 
data set on which the model with the update and the model 
without the update can be compared. Also, the computation 
needed to make a decision on each received update is significant. 
\item {\em Detection via update statistics.} An update is labeled
as bad if it is an outlier with respect to the other updates.
\item {\em Neutralization via aggregation.} Updates are aggregated
using operators that are insensitive to outliers, such as 
the median~\cite{ref37}, the coordinate-wise median~\cite{ref37}, 
or Krum aggregation~\cite{ref34}.
In this way, updates too different from the rest have little
or no influence on the learning process.
\end{itemize}

In our protocol, we want to explicitly detect bad updates in order
to avoid interaction with the malicious peers generating them. Hence,
we discard methods in the third class (neutralization).

Any detection method in the two other classes can be used with our approach, 
including new methods that may appear in the future.
Yet, detection based on model metrics is quite costly and requires
validation data. For this reason,
 in the experimental work we 
have instantiated our implementation with a method based
on update statistics, more specifically 
a distance-based method in line with~\cite{ref36,ref34}.
Given a batch of updates, this method labels as bad an update $U$
if $U$ is much more distant than the rest of updates in the batch
from the batch centroid $C$. 
One possible way to quantify what ``much more distant''
means is to check whether the 
distance between $U$ and $C$ is 
greater than the third quartile (or greater than a
small multiple of the third quartile,
say $1.5$ times)
of the set of distances between updates in the batch and $C$.
 
Protocols~\ref{pro1}, \ref{pro2} and~\ref{pro3} are designed to
incentivize the submission of good updates. Thus, we can state
the following proposition.

\begin{prop}
\label{propsecu}
Provided that the model manager can detect bad updates, the 
rational behavior for generators
and forwardees in Protocol~\ref{pro1} is to submit 
good updates.
\end{prop}

\begin{IEEEproof} See discussion on co-utility for generators and
forwardees in Section~\ref{dis-coutil}.
\end{IEEEproof}

As to collusions of irrationally malicious peers, they 
can only disrupt the learning process if they are sufficiently
large so that the majority of updates received by $M$ are bad ones
{\em and} coordinated in the same direction. Note that
uncoordinated bad updates are likely to cancel each other to some
extent. Such large collusions seem hard to mount for the reasons explained 
in the previous section.

\subsection{Computation and communications overhead}
\label{complex}

Let us compare the computation and communications overhead 
of the proposed method against alternatives based
on homomorphic encryption (HE), which offer a comparable
level of privacy (but cannot detect bad updates,
as argued below).
 
HE has been used in federated 
aggregation mechanisms to prevent the model manager and the
rest of peers in the network from having
access to the individual updates of peers.
In HE-based mechanisms, peers first encrypt their respective updates 
using an additive HE scheme ({\em e.g.} Paillier, \cite{paillier}).
Several protocols have been proposed in the literature 
to aggregate HE updates and decrypt the aggregated HE update.
Let us focus on a protocol that minimizes the number of required
messages and the amount of computation (which is the most challenging
benchmark when comparing with our proposed method): 
(i) assume a sequence of peers 
is defined such that the first peer sends her HE update to the next peer,
who aggregates it with her own HE update and so on;
(ii) after the last peer has aggregated her HE update,
she sends the encrypted update aggregation to the manager,
who can decrypt it to obtain the cleartext update aggregation.
In this protocol, each peer sends only one message per update, just
as in plain federated learning.

Whatever the protocol used, HE-based solutions offer privacy 
(no one other than the peer sees the peer's cleartext
update), but they {\em do not allow} the model manager to detect
bad updates, because the manager does not see the individual
updates. In this respect, HE-based solutions are 
inferior to our proposed method, which 
offers privacy without preventing bad update detection.

Even so, let us compare HE-based 
systems and our system in terms of computational overhead. 
HE-based systems require the peers to
encrypt, using a public-key HE scheme, each individual model parameter
at each training epoch (an update contains
values for all parameters). The authors of~\cite{hecomplex} 
report an encryption
time of $3111.14$ seconds for a model with $900,000$ parameters ($6.87$ MB)
using $3072$-bit Paillier (a key size of $3072$ bits in factorization-based
public-key cryptosystems offers equivalent security to $128$-bit 
symmetric key schemes~\cite{keysizes}).
Expensive modular operations
(with $3072$-bit long moduli in the case of Paillier)
for each model parameter are required to aggregate the update of each
peer.

In contrast, our approach requires each peer to compute 
an encryption of her update using a regular non-homomorphic 
public-key cryptosystem, three 
hashes and one digital signature. 
With the usual digital envelope approach,
regular public-key encryption amounts to encrypting 
a symmetric ({\em e.g.} AES) session key
under the manager's public key, and then encrypting 
the bulk of the update parameters using the 
much faster symmetric cryptosystem under the session key. 
The encryption time of AES on current smartphones 
using AES-128-GCM is around $0.29$
seconds for a model of the same size as reported above\footnote{AES 
performance per
CPU core \url{https://calomel.org/aesni_ssl_performance.html}}, to
be compared with the aforementioned $3111.14$ seconds of HE.
Finally, the model manager just needs to decrypt the received updates
and aggregate them in cleartext as in
plain federated learning mechanisms (this is much 
faster than homomorphic aggregation in ciphertext).

Regarding the communication overhead, we first refer 
to the message expansion incurred
by HE-based mechanisms and our proposal. As stated above, 
HE-based mechanisms require
peers to encrypt each model parameter using an additive HE scheme.
Model parameters are usually $32$-bit floating point values that, 
when encrypted using
Paillier with sufficiently strong keys, become $3072$-bit integers.
This implies an increase in the message size of two orders of magnitude.
The proposal in~\cite{hecomplex} substantially reduces 
the communication requirements, but it is still one order of magnitude
above plain federated learning with cleartext updates.
In our proposal and thanks to the digital
envelope technique, updates are encrypted using a 
symmetric encryption scheme, which does
not expand the plaintext models (save for potential paddings, 
which are negligible
for messages of the size we are considering).
Additionally, our messages include the session key 
encrypted under the
model manager's public key, a triple hash of the model, and a signature.
This additional information increases the size of the message by 
approximately $6.5$ KB with
standard key and hash sizes, which, if we consider the example given before, 
amounts to a
$0.09\%$ increase in the total size of the messages.

Finally, in the HE-based protocol considered 
the number of messages exchanged among participants 
does not increase with respect to plain federated learning, {\em i.e.} 
for each training epoch
there is one broadcast of the global model from the model manager to the peers and one message
from each peer containing her update. In contrast,
our proposal includes a forwarding mechanism, which implies 
that for a forwarding probability $p$ 
every encrypted model hops across an expected number of forwardees   
equal to
\[ (1-p)\sum_{i=1}^\infty ip^{i-1} = \frac{1}{1-p}. \]
For example, if $p=1/2$ there are 2 additional hopping messages
with respect to plain federated learning.
Additionally, if each peer has $m$ accountability managers:
\begin{itemize} 
\item $2m+1$ messages containing one hash of the update and one digital signature of a hash value
are required by the reward protocol;
\item $2m$ messages, of which $m$ are short polling messages
and $m$ contain the signed encrypted update, are
required in the punishment protocol when a peer wishes to avoid 
punishment.  
\end{itemize}
All in all, our approach requires more messages per epoch 
than plain and HE-based federated learning. However, 
whereas the message expansion in our approach is almost
negligible (as the bulk of encryption is symmetric key), 
the HE-based approach increases message length by 
one or two orders of magnitude with respect to plain federated learning.
In particular, if we take say $m=3$ and $p=1/2$, the overall
communications overhead of our approach stays below
that of HE-based federated learning.

In summary, our method achieves {\em much} less computation overhead
and less communication overhead 
than HE-based methods. Add to this 
 performance advantage the functionality advantage: 
our method offers both privacy for peers and detection of bad updates
for the manager, whereas the latter feature is lacking in HE-based
methods. 

\section{Experimental results}
\label{experiments}

In this section we report the results of 
the experiments we conducted to test how 
the reputations of peers evolve over time depending on whether
 they submit good or bad updates.

First, let us explain the expected system behavior.  
If our protocols are well designed, a peer's reputation 
should highly correlate with the probability 
that she generates {\em good} updates. Furthermore, the reputation 
of the peer who {\em submits} an update to the model manager $M$
should also highly correlate with the probability that the 
peer who {\em generated} that update generates good updates. Since 
the submitting peer's reputation is used by $M$ 
to decide on processing or discarding an update, $M$ will only
process a fraction of the received updates. This reduces
 $M$'s overhead related to detection and punishment of {\em bad} updates.

Now, let us go to the actual empirical results.
We bounded the range of reputations between 0 and 1.
Then we built 
a peer-to-peer network with $100$ peers whose 
initial reputations were set to $0$. 
We let the network evolve for $500$ iterations 
(or global training epochs). At each epoch,
the model manager received one update from each peer. 
Thus, the batch size
was $b=100$ and the reward/punishment quantum was $\delta=1/b=0.01$.
We then experimented with two test scenarios, depending 
on the proportion of honest peers:
\begin{itemize}
\item {\em Scenario 1.} Every peer is assigned a random {\em goodness 
probability}
$\pi_g \in_R [0,1]$. With probability $\pi_g$ the peer generates
a good update and with probability $1-\pi_g$ she generates a bad update.
Reputation management is used by peers to decide
on accepting or rejecting a forwarded update and to choose
forwardees.
That is, a peer $P_j$ accepts a forwarded update only if the
requesting peer's reputation is at least $g_j - \alpha$,
where we set $\alpha=0.03$.
In turn, a peer $P_i$ chooses a forwardee based on reputations
as described when explaining the function {\sc Select}()
in Section~\ref{down}.
Additionally, reputation management is also used by the
model manager $M$ to decide on processing or directly
discarding an update submitted by a peer $P_k$. That is, $M$ directly discards
the update with probability
$p_0 (1-\min(g_k/T, 1))$, where we set $p_0=0.5$ and $T=0.5$.
\item {\em Scenario 2.} 90\% of peers always generate good updates
whereas the remaining 10\% have probability $0.2$ 
of generating good updates and probability $0.8$ of 
generating bad updates. 
 Hence we can say
that 90\% of peers have goodness probability $\pi_g=1$ and
10\% of peers have goodness probability $\pi_g=0.2$.
Like in the previous scenario, reputation management is set up by taking
$\alpha=0.03$, $p_0=0.5$ and $T=0.5$.
\end{itemize}

\subsection{Test scenario 1}
\label{tscen2}

In large real federated learning networks with,  
say several thousands or hundreds of thousands of peers
({\em e.g.} smartphones), a small proportion of malicious peers
(even smaller than in Scenario 2) is the most realistic assumption.
Nevertheless, let us study an extreme scenario with
even proportions of good and bad updates.
This will allow us to  
demonstrate that the goodness probability of a peer correlates 
with her reputation and with the reputations of the 
peers submitting her updates. 

Let us assign a random goodness probability in 
the interval $[0,1]$ to each of the 100 peers. 
Thus, on average we can expect peers to generate 
good updates only half of the time. 
Reputations are computed after each of the $500$ global 
training epochs and are used to decide, on the one hand, on update acceptance  and  forwarding (peers accept updates from and forward updates  to  other  peers  depending  on the flexibility parameter $\alpha=0.03$), and on the other hand, on update processing and discarding by the model manager 
(it directly discards updates with probability
$p_0 (1-\min(g_i/T, 1))$, with $p_0=0.5$ and $T=0.5$).

\begin{figure}[!h]
\centering
\includegraphics[width=.5\textwidth]{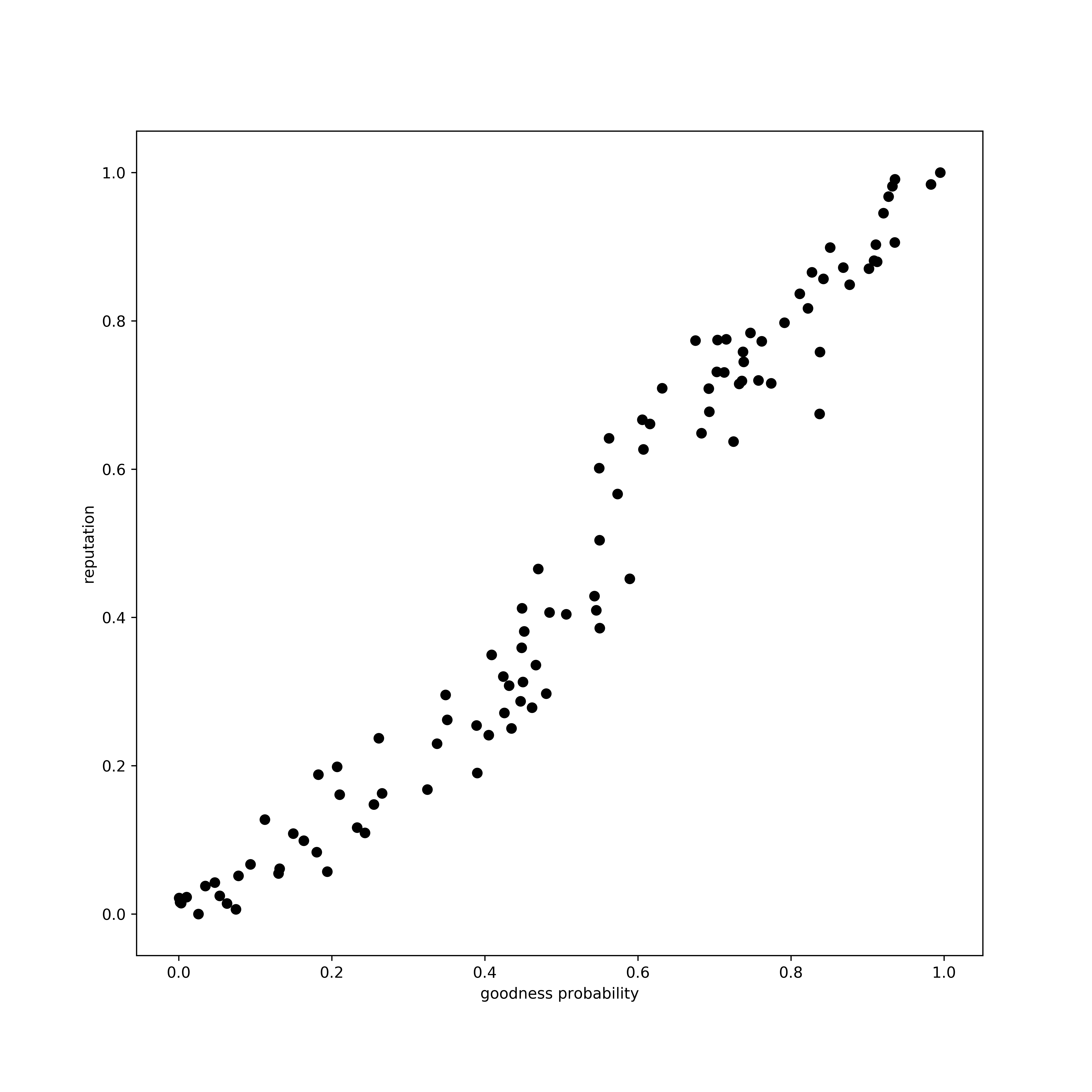}
\caption{Scenario 1. Goodness probability vs reputation 
for each peer. Correlation: 0.977.}
\label{corrscen2}
\end{figure}

Figure~\ref{corrscen2} 
displays the goodness probability versus the
reputation of every peer after the $500$ global training epochs. The goodness probability is represented in the abscissae and
the reputation in the ordinates.
It can be seen that
both the goodness probabilities and their corresponding reputations
 spread over the entire $[0,1]$ range. Furthermore, the peers' goodness probabilities and their reputations are highly correlated (0.977).

\begin{figure}[!h]
\centering
\includegraphics[width=.5\textwidth]{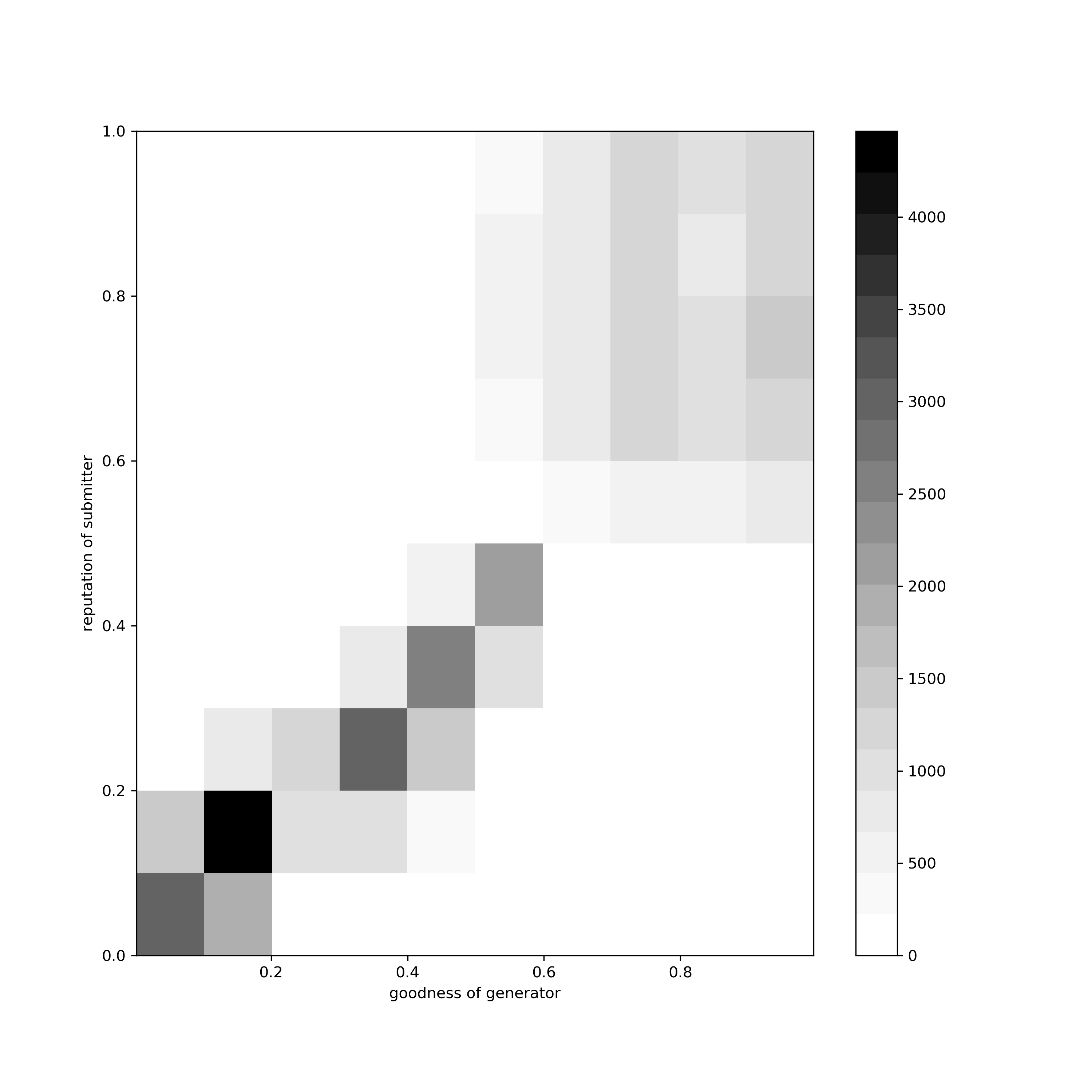}
\caption{Scenario 1. Generating peer's goodness probability
vs submitting peer's reputation, for all updates. The grayscale 
indicates the number of peers in each 2-dimensional interval.
 Correlation: 0.838.}
\label{good_last_peer_scen2}
\end{figure}

Figure~\ref{good_last_peer_scen2} 
displays, for every update during the $500$ global training epochs 
(50,000 updates), the goodness probability of the update 
generating peer versus the
reputation of the submitting peer. 
It can be seen that both values are also highly correlated (0.833).
In fact, this correlation is even higher for peers with reputation below 
$T=0.5$;
for submitting peers with reputations $T=0.5$ or above, the precise
reputation of the submitter is not that relevant, because the model 
manager will process all updates submitted by peers with reputation 
$T$ or above.

\subsection{Test scenario 2}
\label{tscen1}

The previous scenario is highly unlikely in the real world. As said above, in large real federated learning networks a small proportion of malicious peers is the most realistic assumption.

In Scenario 2, a clear majority of 90\% of peers are completely honest (goodness probability $\pi_g=1$), whereas the remaining
10\% have a goodness probability of only $\pi_g=0.2$. Reputations are computed after each epoch and are used to decide, on the one hand, on update acceptance and forwarding, and on the other hand, on update processing and discarding by the 
model manager.

\begin{figure}[!h]
\centering
\includegraphics[width=.5\textwidth]{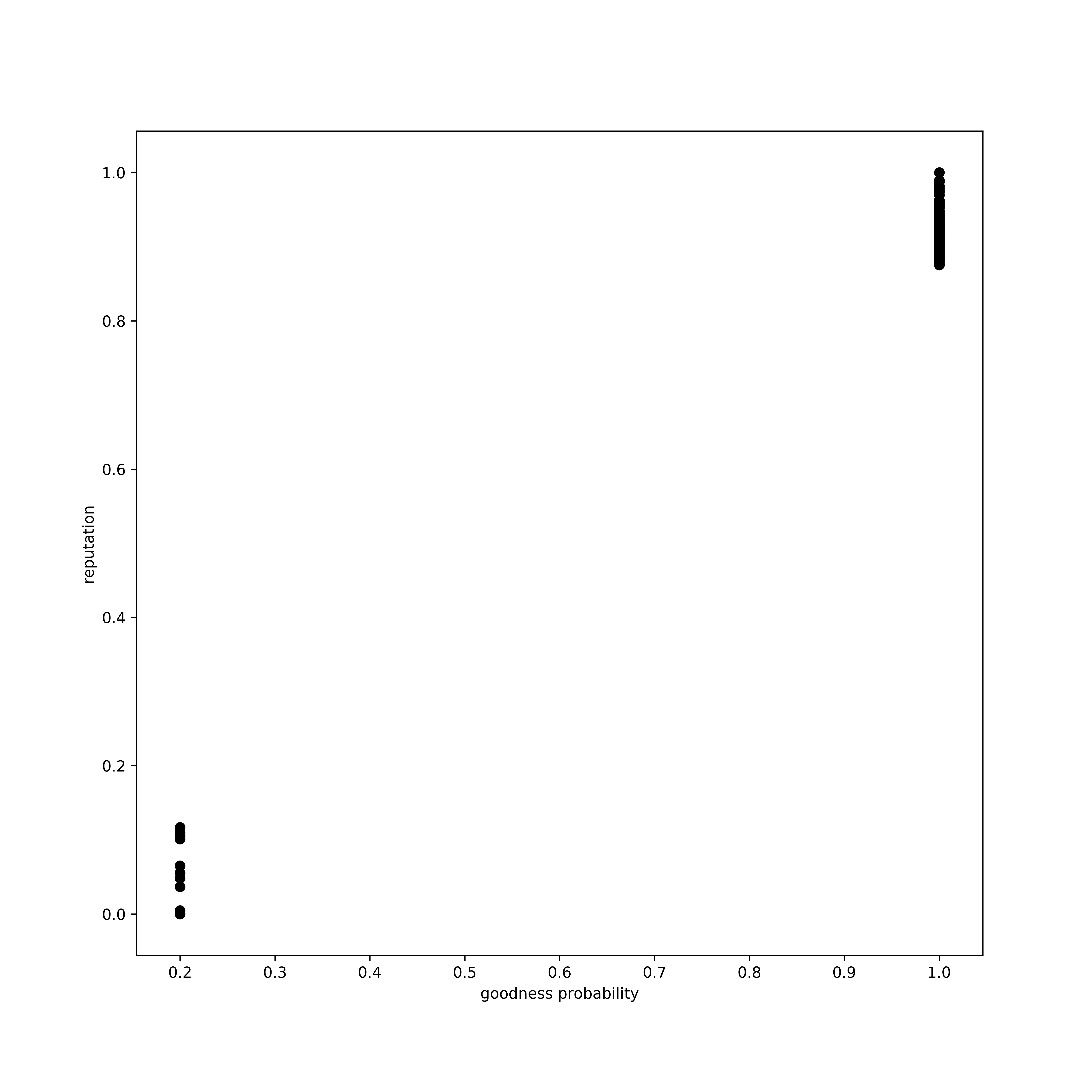}
\caption{Scenario 2. Goodness probability vs reputation 
for each peer. Correlation: 0.998.}
\label{corrscen1}
\end{figure}

Figure~\ref{corrscen1} displays the goodness probability against the
reputation of every peer after $500$ global training epochs. 
Malicious peers (those with $\pi_g=0.2$) are correctly assigned 
low reputations, because most of the updates they generate
are bad and they are punished when their updates reach the model. Besides that, it is hard for such peers to be selected as forwardees of good updates and 
thereby improve their reputation.
On the other side, all honest users (those with $\pi_g=1.0$) achieve high reputation values that correspond to their good behavior. Peers with a reputation 
$T=0.5$ or above are part of a ``community'' whose members improve the 
reputations of each other, by forwarding or submitting their respective updates.

\begin{figure}[h!]
\centering
\includegraphics[width=.5\textwidth]{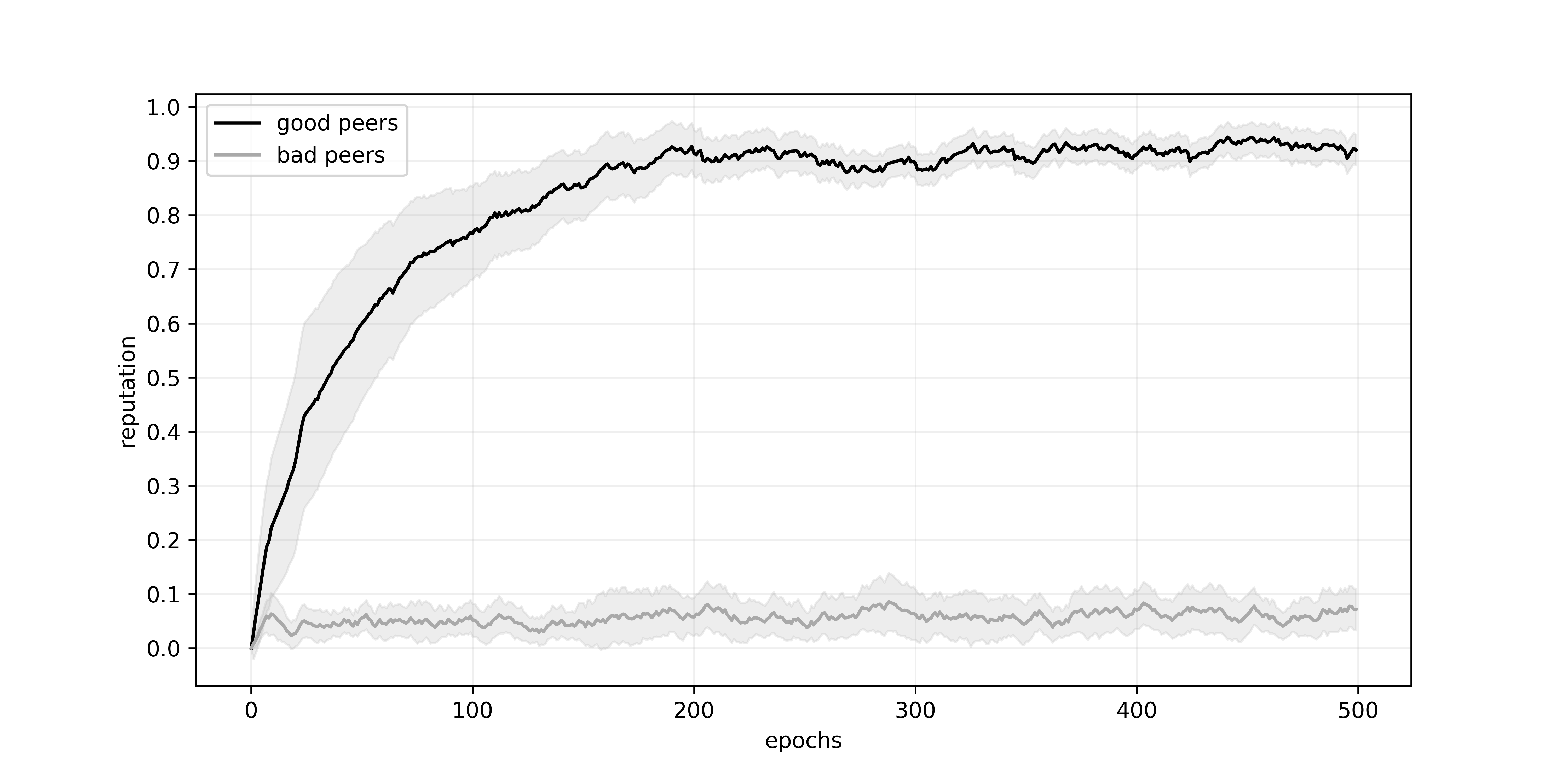}
\caption{Scenario 2. 
Evolution of the average (depicted as a line) and the
standard deviation (depicted as a gray band) of the reputations 
of good peers and bad peers as a function of the epoch.}
\label{evoRepo1}
\end{figure}

The evolution of the reputations of good peers (with $\pi_g=1.0$) and bad peers ($\pi_g=0.2$) is shown in Figure~\ref{evoRepo1}. The average reputations of both types of peers swiftly diverge from the very beginning. 

\begin{figure}[h!]
\centering
\includegraphics[width=.5\textwidth]{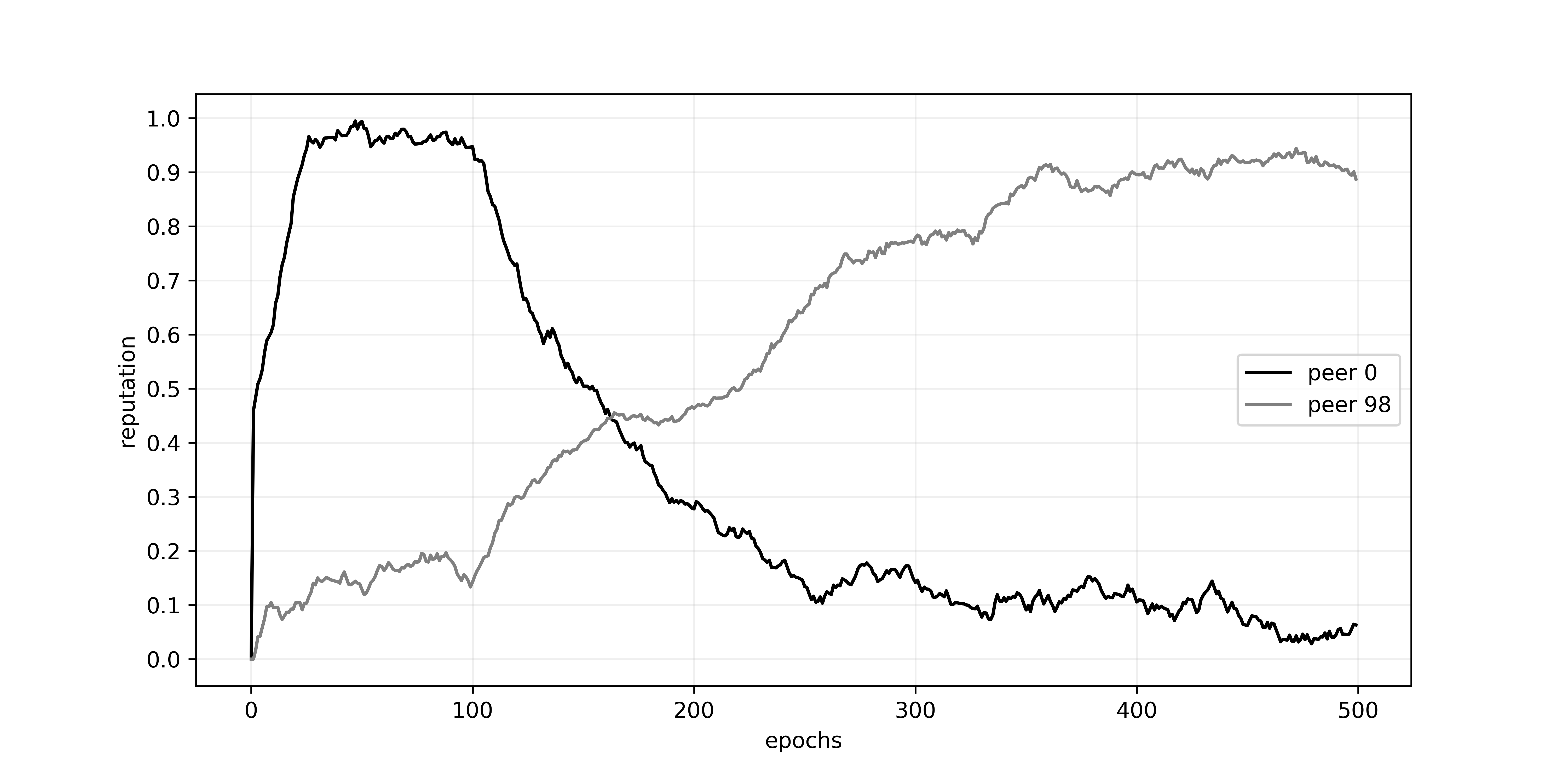}
\caption{Scenario 2. 
Evolution of the reputation of nodes who change their behavior $\pi_g$. 
At epoch 100, peer 0 changes from good to malicious, whereas peer 98 changes
from malicious to good.}
\label{evoRepoChangingGoodness1}
\end{figure}

Figure~\ref{evoRepoChangingGoodness1} shows how reputations evolve when 
peers change their behavior (that is, their $\pi_g$).
In the figure, 
peer 0 is a good peer (with $\pi_g=1.0$) that suddenly changes his 
behavior by setting $\pi_g=0.2$ at epoch 100; from that epoch onwards, 
peer 0 generates bad updates with probability 0.8. 
We can see that his reputation drops fast and stabilizes around the average reputation value of bad peers (see Figure~\ref{evoRepo1}) around epoch 260. This shows that our system reacts suitably when a peer's behavior worsens.

On the other hand, peer 98 in the figure 
represents a malicious peer (with $\pi_g=0.2$) that changes her behavior by setting $\pi_g=1.0$ at epoch 100; from that epoch onwards, peer 98 only 
generates good updates.
 In this case we see that her reputation gradually and slowly increases up to 
roughly the average reputation of good peers (see Figure~\ref{evoRepoChangingGoodness1}) around epoch 360. This shows that not only 
malicious peers, but also newcomers (who have zero initial reputation), 
can effectively reach high reputations if they behave well. However, the amount of effort needed to rise from a low reputation clearly discourages 
malicious peers from performing whitewashing or Sybil attacks.

\begin{figure}[!h]
\centering
\includegraphics[width=.5\textwidth]{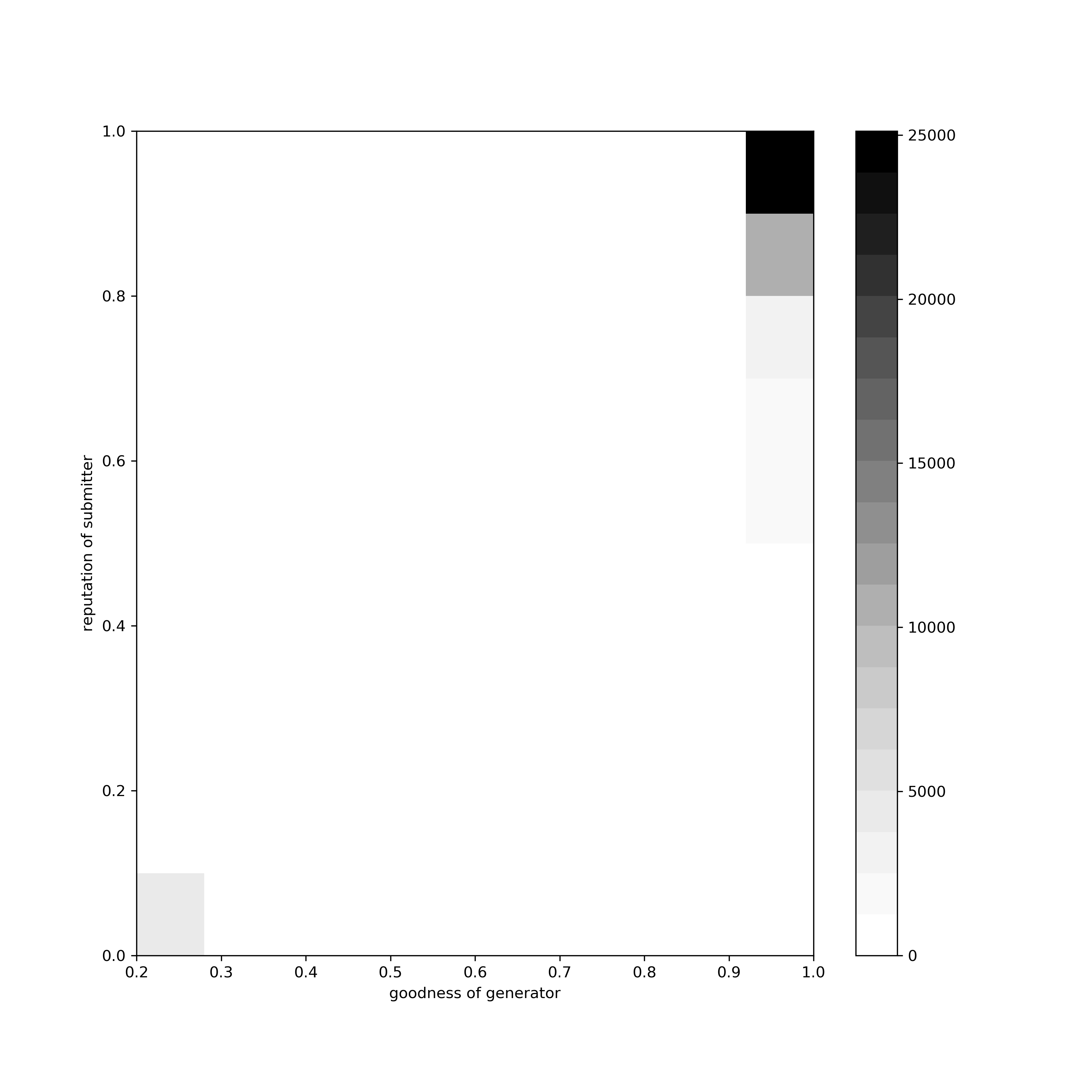}
\caption{Scenario 2. Generating peer's goodness probability vs 
submitting peer's reputation, for all updates.  
The grayscale
indicates the number of peers in each 2-dimensional interval.
Correlation: 0.799.}
\label{good_last_peer_scen1}
\end{figure}

Figure~\ref{good_last_peer_scen1} displays, for every update during the $500$ global training epochs (50,000 updates), the goodness probability of the generator versus the reputation of the submitter. 
Both values are highly correlated (0.799). However, the correlation is higher after the system stabilizes (0.9854 from epoch 100 onwards) and 
all good peers reach high reputations. Initially, reputations have not yet adjusted and hence the updates generated by good peers can be submitted by peers with reputation only slightly above or even slightly below $T$.

\begin{figure}[h!]
\centering
\includegraphics[width=.5\textwidth]{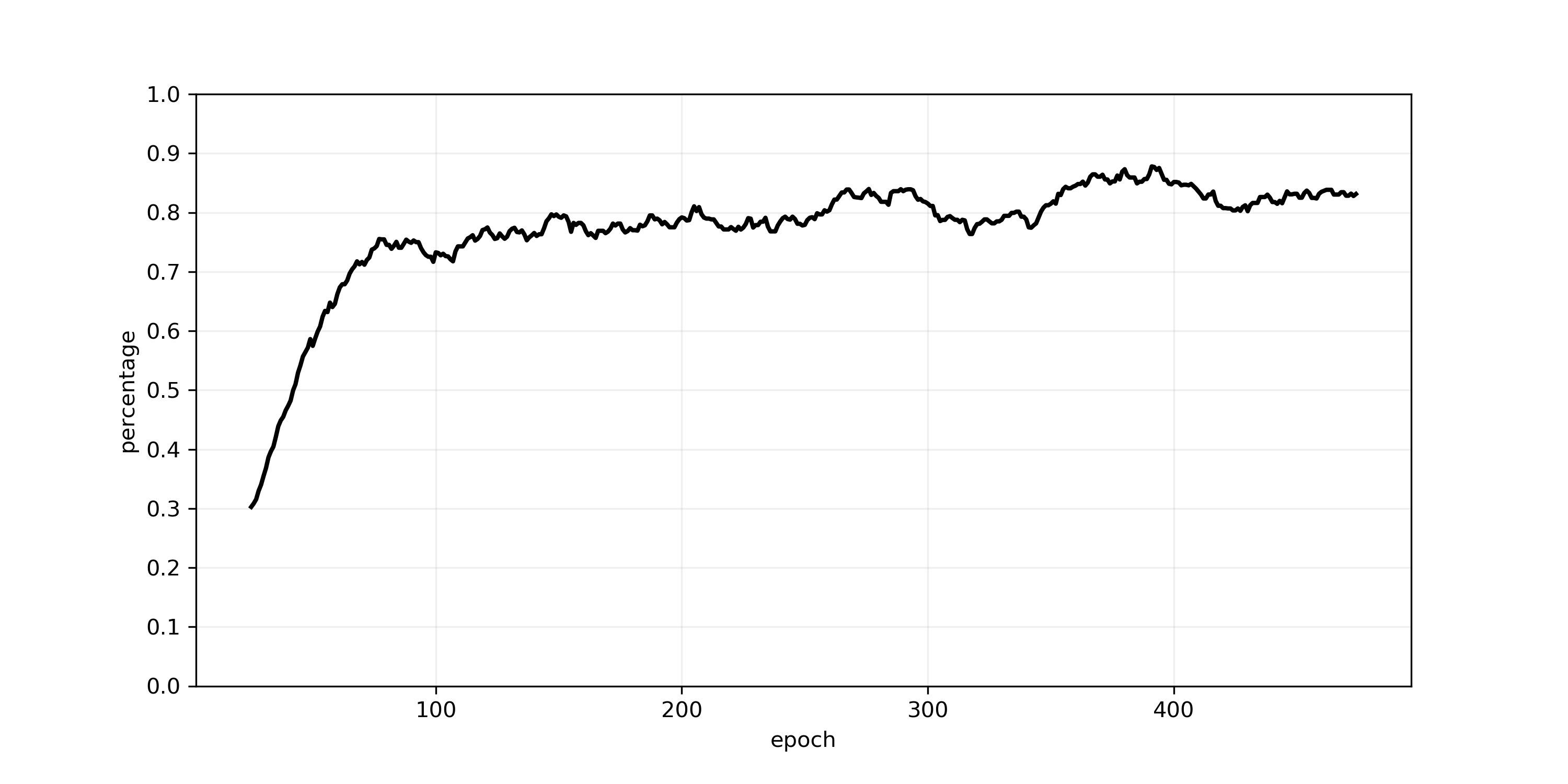}
\caption{Scenario 2. Ratio of bad updates discarded by the 
model manager as a function of the training epoch.}
\label{falseDiscarded1}
\end{figure}

Finally, observe in Figure~\ref{falseDiscarded1}
the effectiveness of making reputation-based decisions to filter out
bad updates. Out of the 50,000 updates generated over the 500 epochs, 
around 46,000 are good, while around 4,000 are bad. Based
on the submitting peer's reputation, the model manager $M$ discards 2,831 updates. 
The figure shows that, when the system stabilizes, on average 
80\% of the updates discarded by $M$ are bad. 
This is the right proportion, because malicious peers do not always
 generate bad updates (they generate bad updates 
with probability $1-\pi_g=0.8$). 

Note that reducing the proportion of bad updates processed by the model manager
is also a good security defense. Indeed, the fewer the bad updates processed by the model manager, the more those bad updates are
likely to stand out as outliers, which will enable $M$ to detect and discard
them. Additionally, fewer bad updates processed by $M$ also mean
less detection overhead for $M$ and, especially, less punishment and
tracing overhead for peers (both normal peers and accountability
managers).

\section{Conclusions and future work}
\label{conclusions}

We have presented protocols to improve privacy and security 
in federated learning while perfectly preserving the model accuracy. 
Our protocols rely on the notion of co-utility,
that is, they are self-enforcing if players are rational. 
We use a decentralized reputation management scheme that is itself co-utile
to incentivize peers to adhere to the prescribed protocols.

In this way, peers do not need to be honest-but-curious {\em per se}: 
as long as they are rational they will behave honestly, 
and even a minority of malicious peers that
do not respond to the same incentives as the other peers 
 can be tolerated.
Confidentiality of the peers' private data is guaranteed by 
the unlinkability of updates: when a peer generates an update,
neither the model manager nor
the other peers can identify the update generator.
This way to provide privacy is superior to the 
state-of-the-art alternatives: 
\begin{itemize}
\item Unlike privacy protection via differential privacy~\cite{ref14},
our protection mechanism does not alter the value of updates and 
hence does not affect the accuracy of the learned model. Furthermore,
our privacy notion based on unlinkability is 
also strong.
\item Unlike privacy protection based on update aggregation,
our solution is compatible with punishing the peers that
generate bad updates. Also, our solution entails less
computational overhead than aggregation based on homomorphic 
encryption.
\end{itemize}

Security, {\em i.e.} protection against bad updates, is pursued
in our approach via reputation.
Whereas state-of-the-art security countermeasures do nothing
to reduce the number of bad updates that are processed by the model
manager, we address this issue in a way to achieve two 
beneficial effects: first, to decrease 
the overhead for the model manager and the
peers related to processing, tracing and punishing bad updates;
and, second, to make the (fewer) bad updates processed by the model manager
more identifiable as outliers.
The design of our protocols also renders whitewashing 
and Sybil attacks ineffective.


An interesting avenue for future research is to harden the proposed
protocols so that they can filter out a greater proportion of 
bad updates in situations where a substantial share 
of the peers are malicious. A possible strategy is for 
the model manager to preventatively reject (without
further examination) any update 
submitted by a peer whose reputation is less than
the average reputation of peers who submitted updates detected
as bad in the past. 
Note that the peer submitting the update
is not the peer having generated it, but as shown 
in the experimental section above, the submitter's and the generator's
reputations are correlated.

Another interesting direction is to incorporate new methods
to detect bad updates that are better suited for 
non-independent and identically distributed (non-IID) private data
than distance-based methods. Most current detection methods
mentioned in Section~\ref{securitysec} 
are ill-suited when the private data of the different peers
follow very different distributions. In fact, the extremely non-IID
case is challenging for the very notion of federated learning 
even if all peers compute their updates honestly: converging 
to an accurate model is more difficult than in the IID case.

\section*{Acknowledgments and disclaimer}
Partial support to this work has been received from the European Commission
(projects H2020-871042 ``SoBigData++'' and H2020-101006879
``MobiDataLab''),
the Government of Catalonia (ICREA Acad\`emia Prizes to 
J. Domingo-Ferrer and D. S\'anchez,
and grant 2017 SGR 705),
and from the Spanish Government (project RTI2018-095094-B-C21
``Consent'' and TIN2016-80250-R ``Sec-MCloud'').
The authors are with the UNESCO Chair in 
Data Privacy, but the views in this paper are their own and are not
necessarily shared by UNESCO.

%





\begin{IEEEbiography}[{\includegraphics[width=1in,height=1.25in,clip,keepaspectratio]{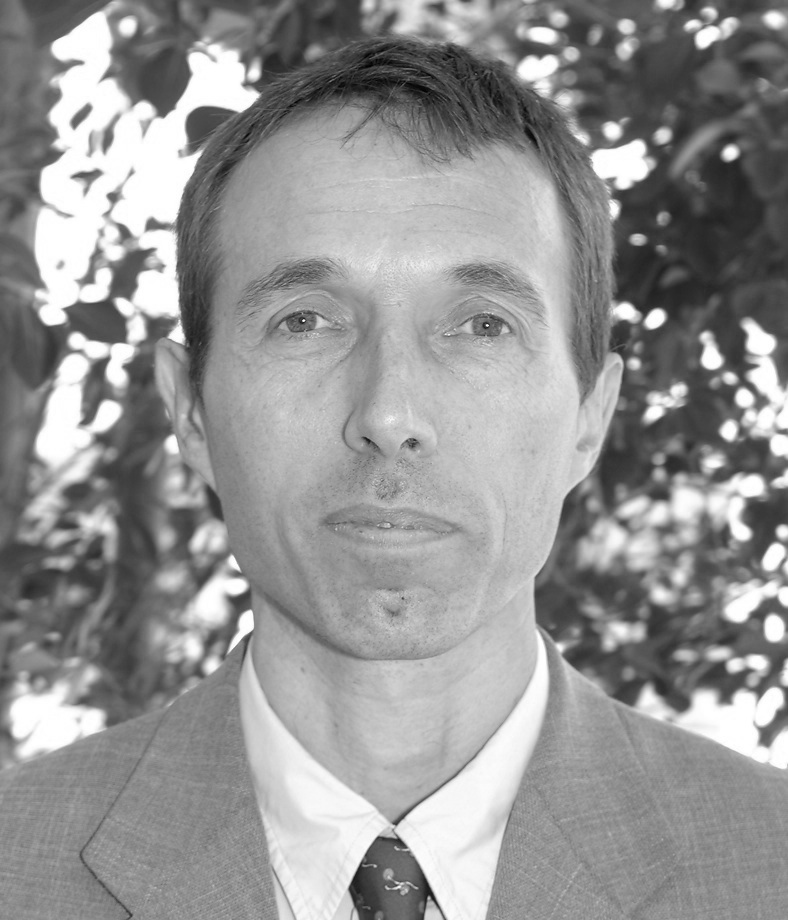}}]{Josep Domingo-Ferrer}
(Fellow, IEEE)
is a distinguished professor of computer science and
an ICREA-Acad\`emia researcher at Universitat Rovira i Virgili,
Tarragona, Catalonia, where he holds the UNESCO Chair in Data Privacy
and leads CYBERCAT. He received the BSc-MSc and
PhD degrees in computer science from
the Autonomous University of Barcelona in 1988 and
1991, respectively. He also holds a BSc-MSc degree in
mathematics.
His research interests are in data privacy, data security and cryptographic
protocols. More information on him can be found
at \url{http://crises-deim.urv.cat/jdomingo}
\end{IEEEbiography}

\begin{IEEEbiography}[{\includegraphics[width=1in,height=1.25in,clip,keepaspectratio]{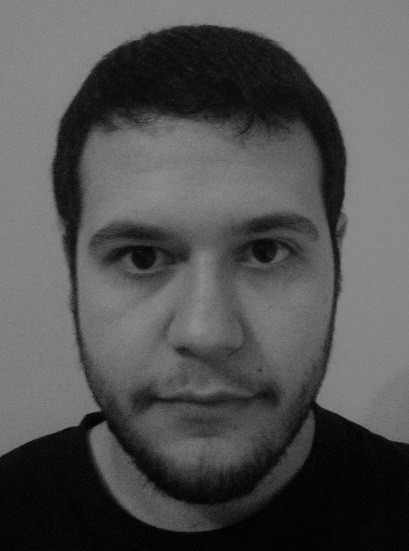}}]{Alberto Blanco-Justicia}
is a postdoctoral researcher at Universitat Rovira i Virgili. He obtained his MSc in computer security in 2013 from Universitat Rovira i Virgili, and his PhD in computer engineering and mathematics of security from the same university in 2017. His research interests are in data privacy, data security, cryptographic protocols, ethically-aligned design and machine learning explainability. He has been involved in several European and national Spanish research projects, as well as technology transfer contracts.
\end{IEEEbiography}

\begin{IEEEbiography}[{\includegraphics[width=1in,height=1.25in,clip,keepaspectratio]{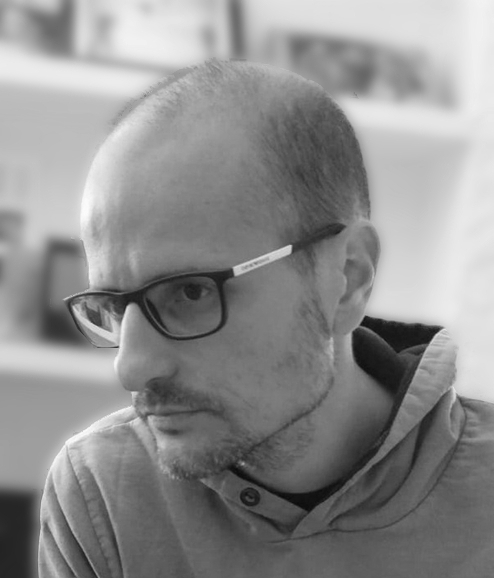}}]{Jes\'us Manj\'on}
is a computer engineer with the UNESCO Chair in Data Privacy and the CRISES
research group at the Department of Computer Engineering
and Mathematics of Universitat Rovira i Virgili, Tarragona, Catalonia.
He received his BSc in computer engineering in 2004 and his
MSc in computer security in 2008. He has participated
in several national and European funded research projects and he is a co-author of
several research publications on security and privacy.
\end{IEEEbiography}

\begin{IEEEbiography}[{\includegraphics[width=1in,height=1.25in,clip,keepaspectratio]{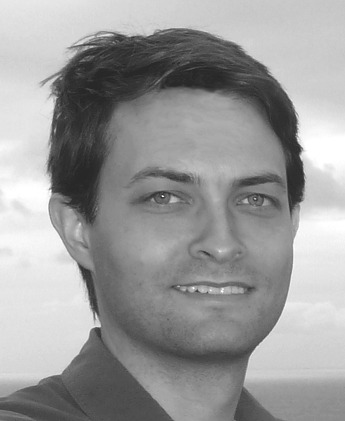}}]{David S\'anchez}
is a Serra Hunter associate professor  
 and an ICREA-Acad\`emia researcher 
at Universitat Rovira i Virgili. He received his PhD in computer science from the Technical University of Catalonia in 2008. He has participated in several national and European-funded research projects and he has authored several papers and conference contributions. His research interests include data semantics, ontologies, data privacy and security.
\end{IEEEbiography}

\end{document}